\documentclass[conference]{IEEEtran}
\IEEEoverridecommandlockouts
\usepackage{amsfonts}
\usepackage{cite,graphicx,amsmath,amsthm}
\usepackage{subcaption}
\usepackage{caption}
\usepackage{fancyhdr}
\usepackage{dsfont}
\usepackage{array,color}
\usepackage{bm}
\usepackage{float}
\usepackage{algorithm}
\usepackage{algpseudocode}
\usepackage{multirow}
\usepackage{booktabs}
\usepackage{multirow}
\usepackage{makecell,balance}
\usepackage{tikz}
\usepackage{amsmath,amsfonts,amssymb,mathrsfs}
\usetikzlibrary{positioning} %为了实现相对位置的设定
\usepackage{xcolor} %为了实现不同的颜色

\begin{document}
	\title{\huge Rate-Splitting and Sum-DoF for the
		$K$-User MISO Broadcast Channel with Mixed CSIT and Order-$(K-1)$ Messages	}
	
	\author{\IEEEauthorblockN{Shuo Zheng$^*$, Tong Zhang$^\dagger$,  Jingfu Li$^\#$, Shuai Wang$^\&$,   Weijie Yuan$^*$,  Gaojie Chen$^\#$,  Rui Wang$^*$}
		$^*$ Department of EEE, Southern University of Science and Technology,
		Shenzhen 518055, China \\
		
		$^\dagger$ Department of  EE, Jinan University, Guangzhou 510632, China \\
		
		$^\&$ Shenzhen Institute of Advanced Technology, Chinese Academy of Sciences, Shenzhen 518055, China \\
		
		$^\#$ 5GIC \& 6GIC, Institute for Communication Systems (ICS), University of Surrey, Guildford GU2 7XH, UK\\
		
		\textsc{Email}: zhengs2021@mail.sustech.edu.cn
		\vspace{-0.4cm}
		%zhangt77@jnu.edu.cn
	}
	\maketitle	
	
	\begin{abstract}
		In this paper, we propose a rate-splitting design and characterize the sum-degrees-of-freedom (DoF) for the $K$-user multiple-input-single-output (MISO) broadcast channel with mixed channel state information at the transmitter (CSIT) and order-$(K-1)$ messages, where mixed CSIT refers to the delayed and imperfect-current CSIT, and order-$(K-1)$ message refers to the message desired by $K-1$ users simultaneously. In particular, for the sum-DoF lower bound, we propose a rate-splitting scheme embedding with retrospective interference alignment. In addition, we propose a matching sum-DoF upper bound via genie signalings and extremal inequality. Opposed to existing works for $K=2$, our results show that the sum-DoF is saturated with CSIT quality when  CSIT quality thresholds are satisfied for $K>2$.
	\end{abstract}

	%\begin{IEEEkeywords}
	%	DoF, mixed CSIT, rate-splitting, $K$-user MISO broadcast channel
	%\end{IEEEkeywords}

	\section{Introduction}	
	
	In practice, communications are always in the presence of imperfect channel state information at the transmitter (CSIT). Usually, the imperfection comes from channel feedback, channel state information (CSI) quantization, and estimation, resulting in CSIT distortion \cite{clerckx2023primer,zhang2022dof,Kangda, Derrick}. A lot of research efforts have been devoted to the study of fundamental multiplexing limits, also known as 
	degrees-of-freedom (DoF), in the presence of imperfect CSIT \cite{Clerckx2,hao2017achievable,Davoodi,zhang2023ratesplitting,Clerckx1,ShengYang,Gou,Yi,Rezaee,Wangzhao,Mohanty,Kerret}. The results of two important works in \cite{hao2017achievable} and \cite{Davoodi} showed that the rate-splitting is an optimal transmission method in achieving sum-DoF of the two-user multiple-input single-output (MISO) broadcast channel with imperfect-current CSIT. A following-up work considering hybrid private and common messages was presented in \cite{zhang2023ratesplitting}. The authors of \cite{Clerckx1} investigated the sum-DoF for the $K$-user MISO broadcast channel with imperfect-current CSIT.
	
	Aside from imperfect-current CSIT, when the wireless channel is temporally correlated and the feedback is delayed, there can be a mix  of delayed and imperfect-current CSIT \cite{ShengYang,Gou,Yi,Rezaee,Wangzhao,Mohanty,Kerret}. In the presence of mixed CSIT,	the DoF region of two-user MISO broadcast channel was characterized in \cite{ShengYang} and \cite{Gou}. In particular, optimal use of delayed and imperfect current CSIT was found in \cite{Gou}. For the two-user broadcast channel and interference channel, the DoF region was derived in \cite{Yi}. For the two-user MIMO interference channel with delayed and imperfect current CSIT, a simplified precoding strategy was proposed in \cite{Rezaee}. The authors of \cite{Wangzhao} obtained the DoF region of two-hop MISO broadcast channel with delayed and imperfect current CSIT. In \cite{Mohanty}, the DoF region of two-user Z MIMO interference channel with delayed and imperfect current CSIT was derived. For the $K$-user MISO broadcast channel with delayed and imperfect current CSIT, the optimal transmission scheme and sum-DoF were obtained in \cite{Kerret}, on the condition that the number of transmit antennas is not less than $K$. Although, the problem of the sum-DoF of the $K$-user MISO broadcast channel with mixed CSIT remains open for general transmit antenna settings.
	
	In this paper, we propose a rate-splitting scheme and characterize the sum-DoF for the $K$-user MISO broadcast channel with mixed CSIT and order-$(K-1)$ messages. We consider that the number of transmit antennas is arbitrary, and the messages are desired by $K-1$ users called order-$(K-1)$ message\footnote{It was shown in \cite{Maddah-Ali} that the DoF of order-$1$ messages, i.e., 
		private messages, can be recursively expressed by that by higher-order messages.}. Specifically, we propose a rate-splitting scheme embedding with retrospective interference alignment, as the  sum-DoF lower bound. Besides, the matching sum-DoF upper bound is derived via genie signalings and extremal inequality. Finally, results reveal that the sum-DoF is saturated with CSIT quality when  CSIT quality thresholds are satisfied for $K>2$.
	%\newpage
	\section{System Model and Definition}
	
	%	\subsection{Three-User MISO Broadcast Channel}
	We consider a $K$-user MISO broadcast channel, where one $M$-antenna transmitter is denoted by $\text{Tx}$ and $K$ single-antenna receivers are denoted by $\text{Rx}_1$, $\text{Rx}_2,...,\text{Rx}_K$, respectively. At the time slot $t$, the received signal at $\text{Rx}_i$ is expressed as
	\begin{equation}
		y_i(t)=\bm{h}_i^H(t)\bm{x}(t)+z_i(t), \,\,\, i=1,...,K, 
	\end{equation}
	where $\bm{h}_i^H(t)\in \mathbb{C}^{1\times M}$ denotes the channel state information (CSI) matrix for $\text{Rx}_i$, $\bm{x}(t)\in \mathbb{C}^{M\times 1}$ denotes the transmitted signal subject to power constraint $\mathbb{E}\left(\Vert \bm{x}(t) \Vert^2\right)\leq P$, and $z_i(t)\sim \mathcal{N}_{\mathbb{C}}(0,1)$ denotes the additive white Gaussian noise (AWGN). For convenience, we further define $\mathcal{H}(t) := [\bm{h}_1(t),\bm{h}_2(t),...,\bm{h}_K(t)]^H \in \mathbb{C}^{K \times M}$ and $\mathcal{H}^n := \{\mathcal{H}(t)\}_{t=1}^n$.
	
	%	\subsection{Delayed CSIT with Temporal Correlation}
	
	At time slot $t$, the historical delayed CSI $\mathcal{H}^{t-1}$ is available at $\text{Tx}$. Based on this, $\text{Tx}$ can infer an imperfect estimate of the current CSI $\widehat{\mathcal{H}}(t):= \{\widehat{\bm{h}}_i(t)\}, i=1,2,...,K$. To summarize, the channel estimate is modeled as
	\begin{equation}
		\bm{h}_i(t)=\widehat{\bm{h}}_i(t)+\widetilde{\bm{h}}_i(t), \,\,\, i=1,...,K,
	\end{equation}
	where the estimation error is denoted by $\widetilde{\bm{h}}_i(t)$. Each element of $\widehat{\bm{h}}_i(t)$ and $\widetilde{\bm{h}}_i(t)$ are characterized by $\mathcal{N}_{\mathbb{C}}(0,1-\sigma^2)$ and $\mathcal{N}_{\mathbb{C}}(0,\sigma^2)$. respectively. Furthermore, we assume that
	%	 $\left(\mathcal{H}^{t-1},\widehat{\mathcal{H}}^{t-1}\right)\rightarrow \widehat{\mathcal{H}}(t) \rightarrow \mathcal{H}(t)$ formulates a Markov chain (i.e., 
	$\mathcal{H}(t)$ is independent of $(\mathcal{H}^{t-1},\widehat{\mathcal{H}}^{t-1})$ when conditioned on $\widehat{\mathcal{H}}(t)$. Besides, all $\text{Rx}$s are assumed to know $\mathcal{H}^t$ and $\widehat{\mathcal{H}}^t$ after the transmission completion at time slot $t$, where $\widehat{\mathcal{H}}^n:= \{\widehat{\mathcal{H}}(t)\}_{t=1}^n$. To quantify, we define \textit{CSIT quality metric} as
	\begin{equation}
		\alpha = -\lim\limits_{P\rightarrow \infty}\frac{\log \sigma^2}{\log P}.  
	\end{equation}
	Under this definition, $\mathbb{E}[\vert \bm{h}^H_i\textbf{p} \vert^2]\sim P^{-\alpha}$, where $\alpha \in [0,1]$, if $\textbf{p}$ is a unit-power zero-forcing (ZF) precoding vector (i.e., $\widehat{\bm{h}}_i^H\textbf{p}=0$). $\alpha=0$ means delayed CSIT and $\alpha \rightarrow \infty$ means perfect CSIT. The DoF is defined below.
	
	%	\subsection{Degree-of-Freedom Characterization}
	The order-$(K-1)$ message is referred to as the message desired by $K-1$ receivers and is needless for only one receiver. Let us denote $\mathcal{W}_{-i}$ as the order-$(K-1)$ message needless for $\text{Rx}_i$ and $R_{-i}=\frac{\log \vert \mathcal{W}_{-i} \vert}{n}$ as the rate. The rate tuple $(R_{-1}(P),...,R_{-K}(P))$ is achievable if the average decoding error probability of each user tends to zero as channel uses $n\rightarrow \infty$. The capacity region $\mathcal{C}_{K-1}(P)$ is defined as the supremum of all achievable rate tuples. Thus, the DoF region $\mathcal{D}$ is defined as
	\begin{align}
		\!\!\!\mathcal{D}=\left\{\begin{array}{l}
			\!\!\!\left\{d_{-1},  ..., d_{-K}\right\} \in \mathbb{R}_{+}^K \mid \\
			\,\,\,\,\,\, \left\{R_{-1}(P), ..., R_{-K}(P)\right\} \in \mathcal{C}_{K-1}(P), \\
			\,\,\,\,\,\, d_{-i}=\lim _{P \rightarrow \infty} \frac{R_{-i}(P)}{\log _2 P}, i=1,..., K.
		\end{array} \right\}.\!\!\!  
	\end{align}
	Accordingly, the sum-DoF is defined as $\sum_{i=1}^K d_{-i}$.
	%	 $\textstyle\sum_{i=1}^K d_{-i}=\lim_{P\rightarrow \infty}\frac{C_{\Sigma}}{\log P}$.
	%	\section{Main Results and Discussion}
	%\label{sec:paper-format}

	\section{Proposed Rate-Splitting Scheme}
	
	In this section, we propose a rate-splitting transmission scheme embedding with retrospective interference alignment. Furthermore, its performance serves as the sum-DoF lower bound of the $K$-user MISO broadcast channel with mixed CSIT and order-$(K-1)$ messages. It is noted that the proposed scheme below considers the situation $M>1$.
	
	\subsection{When $\alpha  \le \frac{1}{K-1}$: Rate-Splitting Embedding with Retrospective Interference Alignment}\label{I}
	
	First of all, we briefly review the proposed scheme. This scheme has two phases. In Phase-I, a part of data symbols are transmitted with random beamforming and a part of data symbols are transmitted with ZF beamforming. In Phase-II, interference is re-constructed and transmitted with random beamforming and a part of data symbols are transmitted with ZF beamforming. Retrospective interference alignment refers to the interference re-construction and transmission, which aligns the interference at receivers. 
	
	Next, the details of this scheme are elaborated.

	\begin{figure}[t]
		\centering
		\includegraphics[width=.85\linewidth]{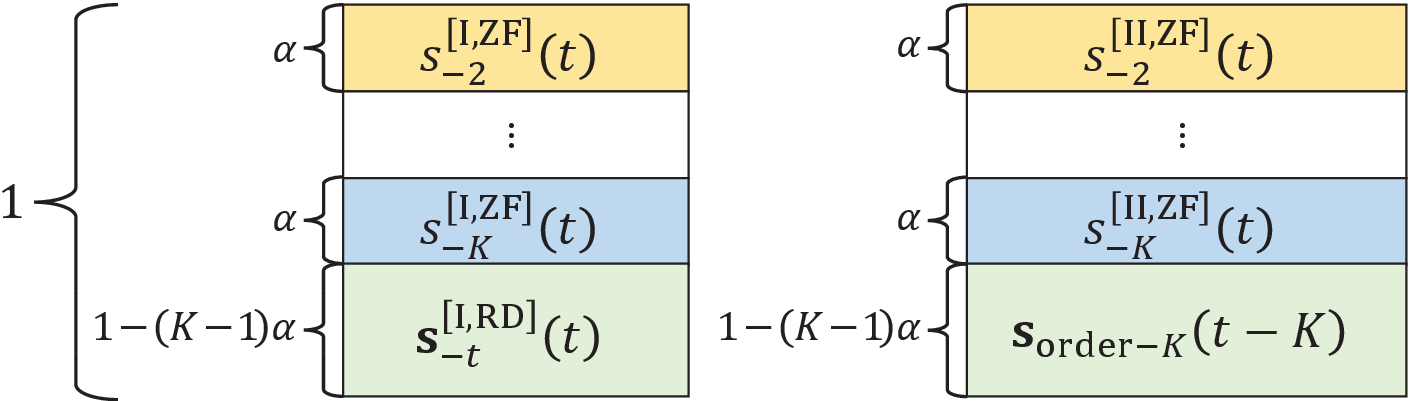}
		\caption{Phase-I and II received signal power level at  $\text{Rx}_1$.} 
		\centering
		\vspace{0.2cm}
		\includegraphics[width=1.5in]{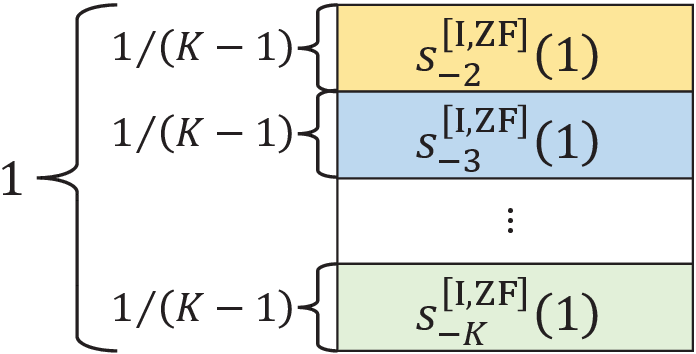}
		\caption{Phase-I received signal power level at  $\text{Rx}_1$.}
		\vspace{-0.5cm}
	\end{figure}

	\subsubsection{Phase-I ($K$ Time Slots)} At the time slot $t=1,2,...,K$,  $\text{Tx}$ transmits data symbols $\textbf{s}_{-1}^{[\text{I,RD}]}(t) \in \mathbb{C}^{\min\{M,2\}},s_{-1}^{[\text{I,ZF}]}(t)\in \mathbb{C},s_{-2}^{[\text{I,ZF}]}(t)\in \mathbb{C},...,s_{-K}^{[\text{I,ZF}]}(t)\in \mathbb{C}$. The transmitted signal at the time slot $t$ is designed as follows:
	\begin{equation}
		\bm{x}(t) =  \textbf{R}(t)\textbf{s}_{-t}^{[\text{I, RD}]}(t) + \sum_{i=1}^K \textbf{p}_i(t) s_{-i}^{[\text{I,ZF}]}(t), 
	\end{equation}
	where $\textbf{R}(t) \in \mathbb{C}^{\min\{M,2\} \times \min\{M,2\}}$ is a unit-power random beamformer whose items are generated randomly, $\textbf{p}_i(t) \in \mathbb{C}^{\min\{M,2\}}$ is a unit-power ZF beamformer that satisfies $\widehat{\bm{h}}_i^H(t)\textbf{p}_i(t) = 0$.   Furthermore,  the power of $\textbf{s}_{-1}^{[\text{I,RD}]}(t)$ is set as $P- P^\alpha$. The power  of $ s_{-i}^{[\text{I,ZF}]}(t)$ is set as $(1/K)P^\alpha$.
	
	As such, the received signal at $\text{Rx}_i$ is given by
	\begin{eqnarray}
		&& \!\!\!\!\!\!\!\!\!\!\!\!\!  y_i(t) = \underbrace{\bm{h}_i^H(t) \textbf{R}(t) \textbf{s}_{-t}^{[\text{I,RD}]}(t)}_{\text{denoted by}\,\eta_i,(1-(K-1)\alpha)\log P\,\, \text{bits}} + 
		\underbrace{\widetilde{\bm{h}}^H_i(t)\textbf{p}_i(t)s_{-i}^{[\text{I,ZF}]}(t)}_{\sim P^{0}}   \nonumber \\
		&&  +	\sum_{j\ne i}\underbrace{\bm{h}^H_i(t)\textbf{p}_j(t)s_{-j}^{[\text{I,ZF}]}(t)}_{\sim P^{\alpha}} +  \underbrace{z_i(t)}_{\sim P^0},\,\,\ \, i=1,...,K. 
	\end{eqnarray}
	
	From the received signal, one can see that only $s_{-i}^{[\text{I,ZF}]}$ term falls into AWGN power level. At the end of this phase, the CSI matrices are returned to $\text{Tx}$. Please refer to Fig. 1 for the received power level at $\text{Rx}_1$.
	From ZF parts of received signals, at each time slot of Phase-I, $K\alpha$ DoF can be achieved.  
	
	\subsubsection{Phase-II ($K-1$ Time Slots)}
	
	At the very beginning of Phase-II, the interference  in Phase-I at $\text{Rx}_i$, i.e., $\bm{h}_i^H(1) \textbf{R}(1) \textbf{s}_{-i}^{[\text{I,RD}]}(1),$ can be reconstructed at $\text{Tx}$. The order-$K$ symbols are thus made by \begin{eqnarray}
		\textbf{s}_\text{order-$K$} :=  \begin{bmatrix}
			\eta_1 + \eta_2 \\
			\eta_2 + \eta_3\\
			...\\
			\eta_{K-1} + \eta_K
		\end{bmatrix} \in \mathbb{C}^{K-1}.  
	\end{eqnarray}
	Besides, data symbols $s_{-i}^{[\text{II,ZF}]}\in \mathbb{C},i=1,2,...,K$ are transmitted in Phase-II. The transmitted signal at the time slot $t=K+1,...,2K-1$ is designed as follows:
	\begin{equation}
		\bm{x}(t) =  \textbf{r}(t) \textbf{s}_\text{order-$K$}(t-K) + \sum_{i=1}^K \textbf{p}_i(t) s_{-i}^{[\text{II,ZF}]}(t),  
	\end{equation}
	where $\textbf{r}(t) \in \mathbb{C}^{\min\{M,2\}}$ is a unit-power random beamformer whose items are generated randomly, $\textbf{p}_i(t) \in \mathbb{C}^{\min\{M,2\}}$ is a unit-power ZF beamformer that satisfies $\widehat{\bm{h}}_i^H(t)\textbf{p}_i(t) = 0$. In addition,  the power of  $\textbf{s}_\text{order-$K$}(t-K)$ is set as $P - P^\alpha$. The power of  $ s_{-i}^{[\text{II,ZF}]}(t)$ is set as $(1/K)P^\alpha$. 
	
	As such, the received signal at $\text{Rx}_i$ is given by 
	\begin{eqnarray}
		&& \!\!\!\!\!\!\!\!\!\!\!\!\!		y_i(t) = \underbrace{\bm{h}^H_i(t)\textbf{r}(t) \textbf{s}_\text{order-$K$}(t-K)}_{(1-(K-1)\alpha)\log P \,\, \text{bits}}   + \underbrace{\widetilde{\bm{h}}^H_i(t)\textbf{p}_i(t)  s_{-i}(t)^{[\text{II,ZF}]}}_{\sim P^{0}}    \nonumber \\ 
		&& + \sum_{j\ne i} \underbrace{\bm{h}^H_i(t)\textbf{p}_j(t)  s_{-j}(t)^{[\text{II,ZF}]}}_{\sim P^\alpha} + \underbrace{z_i(t)}_{\sim P^0},\,\,\, i=1,...,K. 
	\end{eqnarray}
	From the received signal, one can see that only $s_{-i}^{[\text{II,ZF}]}$ term falls into AWGN power level. From ZF parts of received signals, at each time slot of Phase-II, $K\alpha$ DoF can be achieved.
	
	At the end, the decoding of random beamforming data symbols can adopt \cite[Algorithm 1]{ICCC} using backward/forward cancellation. After that, $\frac{2K(1 - (K-1)\alpha)}{2K-1}$ DoF can be achieved.

	\subsection{When $\alpha > \frac{1}{K-1}$: ZF Transmission}
	
	In this case, each receiver has no space  catering random beamforming data symbols, whose decodability is ensured by retrospective interference alignment. This motivates us to design the following pure ZF transmission scheme. 
	
	This scheme only spans $1$ time slot. The transmitter aims to transmit $s_{-1}^{[\text{I,ZF}]}(1)\in \mathbb{C},\cdots,  s_{-K}^{[\text{I,ZF}]}(1) \in \mathbb{C}$. 
	The transmit signal is thus designed as follows:
	\begin{equation}
		\bm{x}(1) =    \sum_{i=1}^K \textbf{p}_i(1) s_{-i}^{[\text{I,ZF}]}(1),  
	\end{equation}
	where $\textbf{p}_i(1) \in \mathbb{C}^{\min\{M,2\}}$ is a unit-power ZF beamformer that satisfies $\widehat{\bm{h}}_i^H(1)\textbf{p}_i(1) = 0$. The power  of $ s_{-i}^{[\text{I,ZF}]}(t)$ is set as  $P^{1/(K-1)}$.  The received power level  
	at $\text{Rx}_i$ is given by
	\begin{align}
		y_i(1)=&\underbrace{\widetilde{\bm{h}}_i^H(1)\textbf{p}_i(1) s_{-i}^{[\text{I,ZF}]}(1)}_{\sim P^0}\nonumber \\ 
		&\hspace{-0.5cm}+\sum\limits_{j\neq i}\underbrace{\bm{h}_i^H(1)\textbf{p}_j(1) s_{-j}^{[\text{I,ZF}]}(1)}_{\sim P^{1/(K-1)}}+\underbrace{z_i(1)}_{\sim P^0}, \,\,\, i=1,...,K, 
	\end{align}
	which is also illustrated in Fig. 2. It can be seen that	
	the transmitted symbols can be decoded at each desired receiver. As such, $\frac{K}{K-1}$ sum-DoF is achieved, as each message attains $1/(K-1)$ achievable DoF.
	
	\subsection{Summary}
	The sum-DoF achieved by the proposed rate-splitting design is summarized as follows:
	\begin{align}
		\sum\limits_{i=1}^K d_{-i} \geq
		\begin{cases} \dfrac{K(1+\frac{\min\{M,2\}-1}{\min\{M,2\}}\alpha)}{K-1+\frac{1}{\min\{M,2\}}}, &\alpha \le \dfrac{1}{K-1},\\
			\dfrac{K}{K-(\min\{M,2\}-1)}, &\alpha > \dfrac{1}{K-1}.
		\end{cases}
	\end{align}
	It is worth mentioning that the proposed scheme is applied when $M>1$. When $M=1$, the sum-DoF of $1$ can be trivially achieved by sending one symbol at each time slot.
	\section{Proposed Sum-DoF Converse}
	
	In this section, we construct the converse from two parts.
	
	\subsection{Part-I: Upper Bound with Perfect CSIT}
	For the $K$-user MISO broadcast channel with order-$(K-1)$ messages, we first provide a  sum-DoF upper bound with perfect CSIT. 
	Since the DoF of $\text{Rx}_i$ is bounded by $1$, 
	we obtain inequalities
	$\sum_{i=1,i\neq k}^{K}d_{-i}\leq 1, \forall i
	$. 
	%Therefore, we have the DoF outer region as follows:
	%\begin{align}
	%\!\!\!\mathcal{D}\subseteq\left\{\!
	%		\{d_{-1}, d_{-2},..., d_{-K}\} \!\in\! \mathbb{R}_{+}^K\ \middle|\ 
	%		\sum_{k=1,k\neq i}^{K}d_{-k}\leq 1, \forall i.\label{trivial outer}\!
	%\right\}\!\!\!\!
	%\end{align}
	%\begin{align}
	%		\sum_{k=1,k\neq i}^{K}d_{-k}\leq 1, \forall i.\label{trivial outer}
	%\end{align}
	By summing these inequalities up, the sum-DoF upper bound is given by $\sum_{i=1}^K d_{-i} \leq K/(K-1)$. When $M=1$, the sum-DoF should be $1$. Hence, the sum-DoF upper bound with perfect CSIT is finally expressed as 
	\begin{align}
		\sum_{i=1}^K d_{-i} \leq\frac{K}{K-(\min\{M,2\}-1)}\label{trivial upper}.
	\end{align}
	
	The results in \eqref{trivial upper} does not take the influence of mixed CSIT into consideration. By considering that, we derive another sum-DoF upper bound in the rest of this section.
	
	\subsection{Part-II: Upper Bound with Mixed CSIT}
	As for genie signalings, a genie provides $\text{Rx}_2$ with messages $\{W_{-2},W_{-3},...,W_{-K}\}$ and $\text{Rx}_1$'s received signals $y_1(m),\forall m\leq t $. For convenience, we denote notations $\bm{Y}_k^n:=\{y_k(t)\}_{t=1}^{n}$, and $W_{[i:j]}:= \{W_i,W_{i-1},...,W_{j}\},i\geq j$.
	
	By employing Fano's inequality and genie signalings, we can upper bound the achievable rate $R_{-1}$ as
	\begin{align}
		&n(R_{-1}-\mathcal{O}(1)) \overset{(a)}{\leq}I(W_{-1}; y_1^n,y_2^n|W_{[-2:-K]},\mathcal{H}^n,\widehat{\mathcal{H}}^n)\nonumber\\
		&\overset{(b)}{=}\sum\limits_{t=1}^n I(W_{-1}; y_1(t),y_2(t)|W_{[-2:-K]},\bm{Y}_1^{t-1}\!,\bm{Y}_2^{t-1}\!,\mathcal{H}^n,\widehat{\mathcal{H}}^n)\nonumber\\
		&\overset{(c)}{=}\sum_{t=1}^n \Big(h( y_1(t),y_2(t)|\bm{U}_1(t),\mathcal{H}(t))\nonumber\\
		&\hspace{2cm} -h(y_1(t),y_2(t)|W_{-1},\bm{U}_1(t),\mathcal{H}(t)\Big),\!\!\! \label{R1}
	\end{align}
	where $\bm{U}_1(t):=\{W_{[-2:-K]},\bm{Y}_{1}^{t-1},\bm{Y}_{2}^{t-1},  \mathcal{H}^{t-1},\widehat{\mathcal{H}}^{t}\}$,
	%$\bm{U}_{1}(t):= \left\{ W_{[-2:-K]},y_{1}^{t-1},y_{2}^{t-1}, \bm{h}^{t-1},\widehat{\bm{h}}^t\right\}$ and 
	(a) holds by using Fano's inequality, (b) holds by using the chain rule of mutual information, and (c) holds because of the definition of mutual information and the fact that $y_1(t),y_2(t)$ are irrelevant to future states given past and current states.  
	
	Similarly, the sum of achievable rates $R_{-k}, k=2,...,K$ can be bounded as
	%	The sum of achievable rates $\text{Rx}_{-i},\forall i=2,3,...,K$ is bounded as
	\begin{align}
		&\sum\limits_{k=2}^K n(R_{-k}-\mathcal{O}(1)) \overset{(a)}{\leq}
		I(W_{[-2:-K]}; y_1^n|\mathcal{H}^n,\widehat{\mathcal{H}}^n)\nonumber\\
		&\overset{(b)}{=}\sum\limits_{t=1}^n I(W_{[-2:-K]}; y_1(t)|\bm{Y}_1^{t-1},\mathcal{H}^n,\widehat{\mathcal{H}}^n)\nonumber\\
		&\overset{(c)}{=}\sum_{t=1}^n \Big(h(y_1(t)|\bm{U}_2(t),\mathcal{H}(t))\nonumber \\&\hspace{2cm}-h(y_1(t)|W_{[-2:-K]},\bm{U}_{2}(t),\mathcal{H}(t))\Big),\label{R23}
	\end{align}
	where $\bm{U}_{2}(t):=\{\bm{Y}_{1}^{t-1}, \mathcal{H}^{t-1},\widehat{\mathcal{H}}^t\}$,
	%	$\bm{U}_{2}(t):= \left\{y_{1}^{t-1}, \bm{h}^{t-1},\widehat{\bm{h}}^t\right\}$, 
	(a) holds by using Fano's inequality, (b) holds by using the chain rule of mutual information, and (c) holds because of the definition of mutual information and the fact that $y_1(t)$ is irrelevant to future states given past and current states. 
	%		\begin{align}
		%			\mathcal{D}\subseteq\left\{\begin{array}{l}
			%				\left\{d_{-1}, d_{-2},..., d_{-K}\right\} \in \mathbb{R}_{+}^K | \\
			%				\dfrac{d_{-1}}{\min\{M,N_1+N_j\}}+\dfrac{\sum_{k=2}^K d_{-k}}{1}\leq 1+\dfrac{\min\{M,N_1+N_j\}-1}{\min\{M,N_1+N_j\}}\alpha,\forall j\neq 1,\\
			%				\dfrac{d_{-i}}{\min\{M,N_j+N_i\}}+\dfrac{\sum_{\substack{k=1,k\neq i}}^Kd_{-k}}{1}  \leq 1+\dfrac{\min\{M,N_j+N_i\}-1}{\min\{M,N_j+N_i\}}\alpha,\forall i \neq 1, j\neq i.
			%			\end{array}\right\} \label{nontrivial outer redundant}
		%		\end{align}\hrule

	In what follows, we upper bound the weighted sum-rate given by \eqref{wRate},
	where (a) holds from \eqref{R1} and \eqref{R23}, and (b) holds since entropy does not increase due to conditioning (i.e. $h(y_1(t)|W_{[-2:-K]},\bm{Y}_2^{t-1},\bm{U}_{2}(t),\mathcal{H}(t))\leq h(y_1(t)|W_{[-2:-K]},\bm{U}_{2}(t),\mathcal{H}(t))$).
	
	%	To further bound the entropy difference term in \eqref{wRate}, we introduce the following lemma.
	
	\begin{figure}[t]
		\centering
		\includegraphics[width=2.25in]{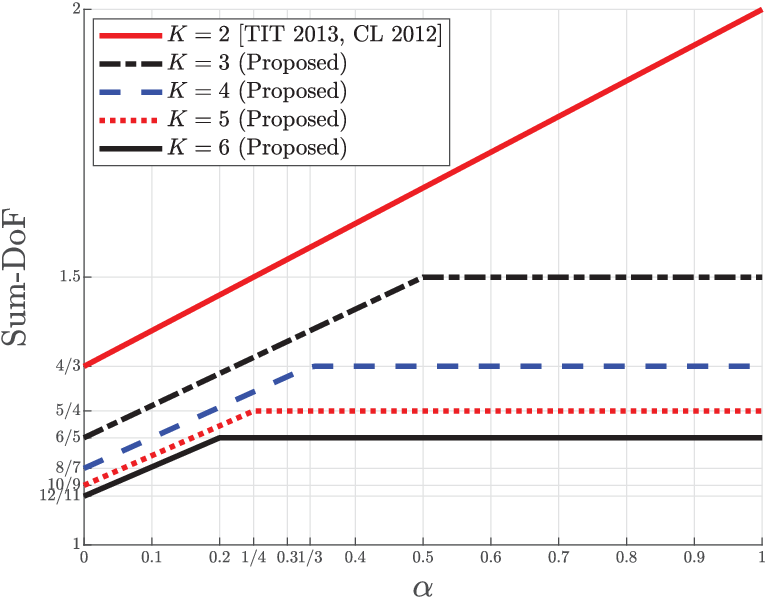}
		\caption{Numerical sum-DoF of the $K$-user MISO broadcast channel with mixed CSIT and order-$(K-1)$ messages.} 
		\vspace{-0.5cm}
	\end{figure}
	Based on Lemma 1 in Appendix A, we have the upper bound of the weighted sum-rate as follows:
	\begin{align*}
		&\frac{n(R_{-1}-\mathcal{O}(1))}{\min\{M,2\}}+\sum_{k=2}^K n(R_{-k}-\mathcal{O}(1))\\
		&\hspace{1.5cm}\leq\! \frac{\min\{M,2\}\!-\!1}{\min\{M,2\}}\alpha n\log P \!+\!  n\log P \!+\! n\mathcal{O}(1).
	\end{align*}
	Therefore, we obtain
	\begin{align*}
		\frac{d_{-1}}{\min\{M,2\}}+\sum_{k=2}^K d_{-k}\leq 1+\frac{\min\{M,2\}-1}{\min\{M,2\}}\alpha.
	\end{align*}
	
	\setcounter{equation}{16}
	By permuting all possible receiver indexes, 
	%	we also have the bounds given by
	%	$
	%		\frac{d_{-i}}{\min\{M, \min\{M,N\}\}}+\!\!\!\!\sum\limits_{\substack{k=1,k\neq i}}^K\!\!\! d_{-k}\leq 1+\frac{\min\{M, \min\{M,N\}\}-1}{\min\{M, \min\{M,N\}\}}\alpha, \forall i=2,3,...,K.
	%	$
	%	Hence, 
	we can obtain inequalities of the outer region in \eqref{nontrivial outer}, which  are shown below.
	\begin{align}
		\dfrac{d_{-i}}{\min\{M,2\}}+\sum_{\substack{k=1,k\neq i}}^Kd_{-k}  \leq 1+\dfrac{\min\{M,2\}-1}{\min\{M,2\}}\alpha,\forall i. \label{nontrivial outer}
	\end{align}

	\subsection{Summary}
	%By combining the results in \eqref{trivial upper} and \eqref{nontrivial upper}, we can obtain the sum-DoF upper bound for $K$-user MISO broadcast channel with mixed CSIT. Note that the bounds in Part-I and Part-II intersects when $K\geq 2$ and $\alpha=1/(K-1)$. Consequently, the complete form of sum-DoF upper bound is given by
	%\begin{figure*}
	%	\hrule
	%	\begin{align}
		%		\mathcal{D}\subseteq\Bigg\{
		%		\left\{d_{-1}, d_{-2},..., d_{-K}\right\} \in \mathbb{R}_{+}^K \ \Bigg|\ \sum_{k=1,k\neq i}^{K}d_{-k}\leq 1, \forall i,\ \ \dfrac{d_{-i}}{\min\{M,2\}}+\sum_{\substack{k=1,k\neq i}}^Kd_{-k}  \leq 1+\dfrac{\min\{M,2\}-1}{\min\{M,2\}}\alpha,\forall i.\Bigg\}\label{nontrivial outer}
		%		%			\dfrac{d_{-i}}{\min\{M,2\}}+\sum_{\substack{k=1,k\neq i}}^Kd_{-k}  \leq 1+\dfrac{\min\{M,2\}-1}{\min\{M,2\}}\alpha,\forall i \neq 1.\right\} \label{nontrivial outer}
		%	\end{align}
	%\end{figure*}

	Based on results in \eqref{trivial upper} and \eqref{nontrivial outer}, we can obtain the sum-DoF upper bound. Note that the bounds in Part-I and Part-II intersects when $K\geq 2$ and $\alpha=1/(K-1)$. Consequently, the complete form of sum-DoF upper bound is given by
	\begin{align}
		\sum\limits_{i=1}^K d_{-i} \leq
		\begin{cases} \dfrac{K(1+\frac{\min\{M,2\}-1}{\min\{M,2\}}\alpha)}{K-1+\frac{1}{\min\{M,2\}}}, &\alpha \le \dfrac{1}{K-1},\\
			\dfrac{K}{K-(\min\{M,2\}-1)}, &\alpha > \dfrac{1}{K-1}.
		\end{cases}
	\end{align}

	\section{Concluding Remarks}
	
	Based on the results of Section-III and -IV, we can conclude that the sum-DoF of the $K$-user MISO broadcast channel with mixed CSIT is given by	
	\begin{equation}
		\sum\limits_{i=1}^K d_{-i} =
		\begin{cases} \dfrac{K(1+\frac{\min\{M,2\}-1}{\min\{M,2\}}\alpha)}{K-1+\frac{1}{\min\{M,2\}}}, &\alpha \le \dfrac{1}{K-1},\\
			\dfrac{K}{K-(\min\{M,2\}-1)}, &\alpha > \dfrac{1}{K-1}.
		\end{cases} \label{Sum}
	\end{equation}
	As illustrated in Fig. 3, our results in \eqref{Sum} show that opposed to existing works \cite{ShengYang} and \cite{Gou} in $K=2$,  the sum-DoF is saturated with CSIT quality when $\alpha > 1/(K-1)$ is satisfied for $K>2$. Besides, we proposed a DoF-optimal rate-splitting scheme with mixed CSIT, which achieves the sum-DoF.

	\section*{Appendix A: Lemma 1}\label{lemma1proof}
	\textbf{Lemma 1.}
	\textit{For the entropy difference term in \eqref{wRate}, we can establish the upper bound  shown below.}
	\begin{eqnarray}
		& \!\!\!\dfrac{1}{\min\{M,2\}} h\left( y_1(t),y_2(t) | \bm{U}_{1}(t), \mathcal{H}(t)\right)\nonumber\\
		&\!\!\!\hspace{2cm}-h\left( y_1(t) | \bm{U}_{1}(t), \mathcal{H}(t)\right)\nonumber\\
		%			& 				\leq \mathbb{E}_{\hat{\bm{h}}(t)} \max_{\substack{\textbf{D} \succeq 0,\\
				%					\operatorname{tr}(\textbf{D}) \leq P}} \mathbb{E}_{\bm{h}(t) | \hat{\bm{h}}(t)}\Big(\dfrac{1}{\min\{M,2\}} \log \operatorname{det}\big(\mathbf{I}+\bm{h}_{[1:2]}(t) \textbf{D}(t) \bm{h}_{[1:2]}^{H}(t)\big)- \dfrac{1}{\min\{M,N_1\}}\log \operatorname{det}\left(\mathbf{I}+\bm{h}_1(t) \textbf{D}(t) \bm{h}^H_1(t)\right)\Big)\label{gaussian} \\
		&\!\!\!\hspace{.6cm}\leq \dfrac{\min\{M,2\}-1}{\min\{M,2\}}\alpha\log P + \mathcal{O}(1).\label{ratebound}\!\!\!  
	\end{eqnarray}

	%\begin{IEEEproof}  
	%	Please refer to Appendix A.
	%\end{IEEEproof}
	\begin{IEEEproof}  To obtain the upper bound of the entropy difference term, we construct an optimization problem to upper bound the term, where the distribution of $\bm{U}_{1}(t)$ and $\bm{x}(t)$ are variables. Based on extremal inequality
		%		in \cite{liu} reveals that optimal $\textbf{x}(t)$ is Gaussian distributed, based on which we have \eqref{gaussian}, where $\bm{h}_{[i:j]}(t):= [\bm{h}^H_i(t),\bm{h}^H_{i+1}(t),...,\bm{h}^H_{j}(t)]^H,i\le j$ and $z_{[i:j]}(t):= [z^H_i(t),z^H_{i+1}(t),...,z^H_j(t)]^H,i\le j$. According to \eqref{gaussian} 
		and \cite[Lemma 3]{Yi}, we have the upper bound in \eqref{ratebound}. 
		The proof of Lemma 1 is elaborated in  \eqref{end},	 
		where $\mathcal{H}_{[i:j]}(t):= [\bm{h}_i(t),\bm{h}_{i+1}(t),...,\bm{h}_{j}(t)]^H,i\le j$,  $\bm{z}_{[i:j]}(t):= [z_i(t),z_{i+1}(t),...,z_j(t)]^T,i\le j$, $\bm{K}$, and $\bm{K}^*$ denote the covariance matrix of $\bm{x}(t)$ and the optimal $\bm{K}$, respectively. The reasons of critical steps are follows:
		(a) The maximization is moved inside the expectation and the resultant value is not less than the original one;
		(b) $\mathcal{H}(t)$ is independent of $(\mathcal{H}^{t-1},\widehat{\mathcal{H}}^{t-1})$ if conditioned on $\widehat{\mathcal{H}}(t)$;
		(c) The maximization is divided into two parts (i.e., trace constraint and covariance matrix constraint);
		(d) The optimal $\bm{x}(t)$ should be Gaussian distributed, which is derived via extremal inequality \cite{liu};
		(e) $\bm{K}^*$ always satisfies $\bm{K}^*\succeq 0, \operatorname{tr}(\bm{K}^*)\leq P$ and inner expectation only relates to  $\widehat{\mathcal{H}}(t)$, where $\operatorname{tr}(\cdot)$ denotes the trace;
		(f) This inequality is obtained from \cite[Lemma 3]{Yi}.
	\end{IEEEproof}
	
	\begin{figure*}
		\vspace{-1cm}
		\begin{align}
			&\frac{n(R_{-1}-\mathcal{O}(1))}{\min\{M,2\}}+\sum_{k=2}^K n(R_{-k}-\mathcal{O}(1))\notag\\
			& \overset{(a)}{\leq} \sum_{t=1}^n \left\{ \frac{1}{\min\{M,2\}}h(y_1(t),y_2(t)| \bm{U}_{1}(t),\mathcal{H}(t))
			-\frac{1}{\min\{M,2\}}h(y_1(t),y_2(t)|W_{-1},\bm{U}_{1}(t),\mathcal{H}(t))\right.\nonumber\\
			&\hspace{1.5cm} +h(y_1(t)| \bm{U}_{2}(t),\mathcal{H}(t))
			-h(y_1(t)|W_{[-2:-K]},\bm{U}_{2}(t),\mathcal{H}(t))\bigg\}\nonumber\\
			&= \sum_{t=1}^n \left\{ \left[\frac{1}{\min\{M,2\}}h(y_1(t),y_2(t)| \bm{U}_{1}(t),\mathcal{H}(t))-h(y_1(t)|W_{[-2:-K]},\bm{U}_{2}(t),\mathcal{H}(t))\right]\right.
			\nonumber\\
			&\hspace{3.5cm} 
			-\underbrace{\frac{1}{\min\{M,2\}}h(y_1(t),y_2(t)|W_{-1},\bm{U}_{1}(t),\mathcal{H}(t))}_{\mathcal{O}(1)} +\underbrace{\vphantom{\frac{1}{\min\{M, N_{1,2}\}}}h(y_1(t)| \bm{U}_{2}(t),\mathcal{H}(t))}_{\leq\log P}\bigg\}\nonumber\\
			&\overset{(b)}{\leq} \sum_{t=1}^n \left\{ \frac{1}{\min\{M,2\}}h(y_1(t),y_2(t)| \bm{U}_{1}(t),\mathcal{H}(t))-h(y_1(t)|W_{[-2:-K]},\bm{Y}_2^{t-1},\bm{U}_{2}(t),\mathcal{H}(t))+\log P+ \mathcal{O}(1)
			\right\}	\nonumber\\
			&= \sum_{t=1}^n \left\{ \frac{1}{\min\{M,2\}}h(y_1(t),y_2(t)| \bm{U}_{1}(t),\mathcal{H}(t))-h(y_1(t)|\bm{U}_{1}(t),\mathcal{H}(t))+\log P+\mathcal{O}(1)
			\right\}	.\label{wRate}\tag{16}
			%	& =\sum_{t=1}^n\left\{ \frac{1}{2}h(y_1(t),y_2(t)| \bm{U}_{1}(t),\mathcal{H}(t))-h(y_1(t)|\bm{U}_{1}(t),\mathcal{H}(t))\right\}+n\log P +n \mathcal{O}(1)\label{wRate}
		\end{align}\vspace{-0.1cm}\hrule\vspace{-0.1cm}
		\begin{align}
			& \frac{1}{\min\{M,2\}}h(y_1(t),y_2(t)| \bm{U}_{1}(t),\mathcal{H}(t))-h(y_1(t)|\bm{U}_{1}(t),\mathcal{H}(t))\nonumber \\
			& \leq \max _{p\left(\bm{U}_{1}(t)\right), p\left(\bm{x}(t) | \bm{U}_{1}(t)\right)}\left(\frac{1}{\min\{M,2\}}h(y_1(t),y_2(t)| \bm{U}_{1}(t),\mathcal{H}(t))-h(y_1(t)|\bm{U}_{1}(t),\mathcal{H}(t)) \right)\nonumber\\
			& \overset{(a)}{\leq} \max_{p\left(\bm{U}_{1}(t)\right)} \mathbb{E}_{\bm{U}_{1}(t)} \max _{p\left(\bm{x}(t) | \bm{U}_{1}(t)\right)}\left(\frac{1}{\min\{M,2\}}h(y_1(t),y_2(t)| \bm{U}_{1}(t),\mathcal{H}(t))-h(y_1(t)|\bm{U}_{1}(t),\mathcal{H}(t))\right) \nonumber\\
			& =\max _{p\left(\bm{U}_{1}(t)\right)} \mathbb{E}_{\bm{U}_{1}(t)} \max _{p\left(\bm{x}(t) | \bm{U}_{1}(t)\right)} \mathbb{E}_{\mathcal{H}(t) | \bm{U}_{1}(t)}\left(\frac{1}{\min\{M,2\}}h(y_1(t),y_2(t)| \bm{U}_{1}(t),\mathcal{H}(t))-h(y_1(t)|\bm{U}_{1}(t),\mathcal{H}(t))\right) \nonumber\\
			& \overset{(b)}{=}\max _{p\left(\bm{U}_{1}(t)\right)} \mathbb{E}_{\bm{U}_{1}(t)} \max _{p\left(\bm{x}(t) | \bm{U}_{1}(t)\right)} \mathbb{E}_{\mathcal{H}(t) | \widehat{\mathcal{H}}(t)}\left(\frac{1}{\min\{M,2\}}h(\mathcal{H}_{[1:2]}(t) \bm{x}(t)+\bm{z}_{[1:2]}(t) | \bm{U}_{1}(t)) -h(\bm{h}^H_1(t) \bm{x}(t)+z_{1}(t) | \bm{U}_{1}(t))\right) \nonumber\\
			& \overset{(c)}{=}\max _{p\left(\bm{U}_{1}(t)\right)} \mathbb{E}_{\bm{U}_{1}(t)} \max _{\substack{\bm{D} \succeq 0 \\
					\operatorname{tr}(\bm{D}) \leq P}} \max _{\substack{p\left(\bm{x}(t) | \bm{U}_{1}(t)\right) \\
					\operatorname{cov}\left(\bm{x}(t) | \bm{U}_{1}(t)\right) \preceq \bm{D}}} \mathbb{E}_{\mathcal{H}(t) | \widehat{\mathcal{H}}(t)}\left(\frac{h(\mathcal{H}_{[1:2]}(t) \bm{x}(t)+\bm{z}_{[1:2]}(t) | \bm{U}_{1}(t))}{\min\{M,2\}}-h(\bm{h}^H_1(t) \bm{x}(t)+z_{1}(t) | \bm{U}_{1}(t))\right)\nonumber\\
			& \overset{(d)}{\leq} \max _{p\left(\bm{U}_{1}(t)\right)} \mathbb{E}_{\bm{U}_{1}(t)} \max _{\substack{\bm{D} \succeq 0 \\
					\operatorname{tr}(\bm{D}) \leq P}} \max _{\bm{K}(t) \preceq \bm{D}} \mathbb{E}_{\mathcal{H}(t) | \widehat{\mathcal{H}}(t)}\left(\frac{1}{\min\{M,2\}} \log \operatorname{det}\!\left(\mathbf{I}\!+\!\mathcal{H}_{[1:2]}(t) \bm{K}(t) \mathcal{H}_{[1:2]}^{H}(t)\right)\!-\! \log\left(1\!+\!\bm{h}^H_1(t) \bm{K}(t) \bm{h}_1(t)\right)\right) \nonumber\\
			& =\max _{p\left(\bm{U}_{1}(t)\right)} \mathbb{E}_{\bm{U}_{1}(t)} \!\max _{\substack{\bm{D} \succeq 0 \\
					\operatorname{tr}(\bm{D}) \leq P}}\! \mathbb{E}_{\mathcal{H}(t) | \widehat{\mathcal{H}}(t)}\left(\frac{1}{\min\{M,2\}} \log \operatorname{det}\left(\mathbf{I}+\mathcal{H}_{[1:2]}(t) \bm{K}^*(t) \mathcal{H}_{[1:2]}^{H}(t)\right)-\log \left(1+\bm{h}^H_1(t) \bm{K}^*(t) \bm{h}_1(t)\right)\right)\nonumber \\
			& \overset{(e)}{\leq} \mathbb{E}_{\widehat{\mathcal{H}}(t)} \max _{\substack{\bm{D} \succeq 0 \\
					\operatorname{tr}(\bm{D}) \leq P}} \mathbb{E}_{\mathcal{H}(t) | \widehat{\mathcal{H}}(t)}\left(\frac{1}{\min\{M,2\}} \log \operatorname{det}\left(\mathbf{I}+\mathcal{H}_{[1:2]}(t) \bm{D}(t) \mathcal{H}_{[1:2]}^{H}(t)\right)- \log \left(1+\bm{h}^H_1(t) \bm{D}(t) \bm{h}_1(t)\right)\right)\nonumber \\
			& \overset{(f)}{\leq} \frac{\min\{M,2\}-1}{\min\{M,2\}}\alpha\log P +  \mathcal{O}(1). \label{end}  
		\end{align}
		\vspace{-0.2cm}\hrule
		\balance
		\vspace{-0.6cm}
	\end{figure*}
	
	\bibliographystyle{IEEEtran}	
	\bibliography{VTC-Fall-HK}

\begin{thebibliography}{10}
\providecommand{\url}[1]{#1}
\csname url@samestyle\endcsname
\providecommand{\newblock}{\relax}
\providecommand{\bibinfo}[2]{#2}
\providecommand{\BIBentrySTDinterwordspacing}{\spaceskip=0pt\relax}
\providecommand{\BIBentryALTinterwordstretchfactor}{4}
\providecommand{\BIBentryALTinterwordspacing}{\spaceskip=\fontdimen2\font plus
\BIBentryALTinterwordstretchfactor\fontdimen3\font minus
  \fontdimen4\font\relax}
\providecommand{\BIBforeignlanguage}[2]{{%
\expandafter\ifx\csname l@#1\endcsname\relax
\typeout{** WARNING: IEEEtran.bst: No hyphenation pattern has been}%
\typeout{** loaded for the language `#1'. Using the pattern for}%
\typeout{** the default language instead.}%
\else
\language=\csname l@#1\endcsname
\fi
#2}}
\providecommand{\BIBdecl}{\relax}
\BIBdecl

\bibitem{clerckx2023primer}
B.~Clerckx, Y.~Mao, E.~A. Jorswieck, J.~Yuan, D.~J. Love, E.~Erkip, and
  D.~Niyato, ``A primer on rate-splitting multiple access: Tutorial, myths, and
  frequently asked questions,'' \emph{IEEE J. Sel. Areas Commun.}, vol.~41,
  no.~5, pp. 1265--1308, May 2023.

\bibitem{zhang2022dof}
T.~Zhang, ``The {DoF} region of two‐user {MIMO} broadcast channel with
  delayed imperfect‐quality {CSIT},'' \emph{Electron. Lett.}, vol.~58, pp.
  432--435, 2022.

\bibitem{Kangda}
K.~Zhi, C.~Pan, H.~Ren, K.~Wang, M.~Elkashlan, M.~D. Renzo, R.~Schober, H.~V.
  Poor, J.~Wang, and L.~Hanzo, ``Two-timescale design for reconfigurable
  intelligent surface-aided massive {MIMO} systems with imperfect {CSI},''
  \emph{IEEE Trans. Inf. Theory}, vol.~69, no.~5, pp. 3001--3033, May 2023.

\bibitem{Derrick}
H.~Fu, S.~Feng, W.~Tang, and D.~W.~K. Ng, ``Robust secure beamforming design
  for two-user downlink {MISO} rate-splitting systems,'' \emph{IEEE Trans.
  Wireless Commun.}, vol.~19, no.~12, pp. 8351--8365, Dec 2020.

\bibitem{Clerckx2}
Y.~Mao, O.~Dizdar, B.~Clerckx, R.~Schober, P.~Popovski, and H.~V. Poor,
  ``Rate-splitting multiple access: Fundamentals, survey, and future research
  trends,'' \emph{IEEE Commun. Surveys Tuts.}, vol.~24, no.~4, pp. 2073--2126,
  Fourthquarter 2022.

\bibitem{hao2017achievable}
C.~Hao, B.~Rassouli, and B.~Clerckx, ``Achievable {DoF} regions of {MIMO}
  networks with imperfect {CSIT},'' \emph{IEEE Trans. Inf. Theory}, vol.~63,
  no.~10, pp. 6587--6606, Oct 2017.

\bibitem{Davoodi}
A.~G. Davoodi and S.~A. Jafar, ``Degrees of freedom region of the {(M, N1, N2)}
  {MIMO} broadcast channel with partial {CSIT}: An application of sum-set
  inequalities,'' in \emph{Proc. IEEE Int. Symp. Inf. Theory (ISIT)}, July
  2019, pp. 1637--1641.

\bibitem{zhang2023ratesplitting}
T.~Zhang, Y.~Zhuang, G.~Chen, S.~Wang, B.~Lv, R.~Wang, and P.~Xiao,
  ``Rate-splitting with hybrid messages: {DoF} analysis of the two-user {MIMO}
  broadcast channel with imperfect {CSIT},'' \emph{arXiv preprint}, 2023.

\bibitem{Clerckx1}
H.~Joudeh and B.~Clerckx, ``Sum-rate maximization for linearly precoded
  downlink multiuser {MISO} systems with partial {CSIT}: A rate-splitting
  approach,'' \emph{IEEE Trans. Commun.}, vol.~64, no.~11, pp. 4847--4861, Nov
  2016.

\bibitem{ShengYang}
S.~Yang, M.~Kobayashi, D.~Gesbert, and X.~Yi, ``Degrees of freedom of time
  correlated {MISO} broadcast channel with delayed {CSIT},'' \emph{IEEE Trans.
  Inf. Theory}, vol.~59, no.~1, pp. 315--328, Jan 2013.

\bibitem{Gou}
T.~Gou and S.~A. Jafar, ``Optimal use of current and outdated channel state
  information: Degrees of freedom of the {MISO} {BC} with mixed {CSIT},''
  \emph{IEEE Commun. Lett.}, vol.~16, no.~7, pp. 1084--1087, July 2012.

\bibitem{Yi}
X.~Yi, S.~Yang, D.~Gesbert, and M.~Kobayashi, ``The degrees of freedom region
  of temporally correlated {MIMO} networks with delayed {CSIT},'' \emph{IEEE
  Trans. Inf. Theory}, vol.~60, no.~1, pp. 494--514, Jan 2014.

\bibitem{Rezaee}
M.~Rezaee and P.~J. Schreier, ``A degrees-of-freedom-achieving scheme for the
  temporally correlated {MIMO} interference channel with delayed {CSIT},''
  \emph{IEEE Trans. Wireless Commun.}, vol.~17, no.~8, pp. 5397--5408, Aug
  2018.

\bibitem{Wangzhao}
Z.~Wang, M.~Xiao, C.~Wang, and M.~Skoglund, ``Degrees of freedom of two-hop
  {MISO} broadcast networks with mixed {CSIT},'' \emph{IEEE Trans. Wireless
  Commun.}, vol.~13, no.~12, pp. 6982--6995, Dec 2014.

\bibitem{Mohanty}
K.~Mohanty and M.~K. Varanasi, ``Degrees of freedom region of the {MIMO}
  {Z}-interference channel with mixed {CSIT},'' \emph{IEEE Commun. Lett.},
  vol.~20, no.~12, pp. 2422--2425, Dec 2016.

\bibitem{Kerret}
P.~de~Kerret, D.~Gesbert, J.~Zhang, and P.~Elia, ``Optimal {DoF} of the
  {K}-user broadcast channel with delayed and imperfect current {CSIT},''
  \emph{IEEE Trans. Inf. Theory}, vol.~66, no.~11, pp. 7056--7066, 2020.

\bibitem{Maddah-Ali}
M.~A. Maddah-Ali and D.~Tse, ``Completely stale transmitter channel state
  information is still very useful,'' \emph{IEEE Trans. Inf. Theory}, vol.~58,
  no.~7, pp. 4418--4431, July 2012.

\bibitem{ICCC}
T.~Zhang, S.~Wang, T.~Wang, and R.~Wang, ``The {DoF} region of order-({K} - 1)
  messages for the {K}-user {MIMO} broadcast channel with delayed {CSIT},'' in
  \emph{Proc. IEEE/CIC Int. Conf. Commun. China (ICCC)}, July 2021, pp.
  688--693.

\bibitem{liu}
T.~Liu and P.~Viswanath, ``An extremal inequality motivated by multiterminal
  information-theoretic problems,'' \emph{IEEE Trans. Inf. Theory}, vol.~53,
  no.~5, pp. 1839--1851, May 2007.

\end{thebibliography}
\end{document}